\documentclass{article}
\usepackage[T1]{fontenc} 
\usepackage[utf8]{inputenc} 
\usepackage{ismir,amsmath,cite,url}
\usepackage{graphicx}

\usepackage{color}

\usepackage[firstpage]{draftwatermark}
\definecolor{lightgray}{rgb}{0.9,0.9,0.9}
\definecolor{darkgray}{rgb}{0.4,0.4,0.4}
\SetWatermarkFontSize{12pt}
\SetWatermarkScale{1.1}
\SetWatermarkAngle{90}
\SetWatermarkHorCenter{202mm}
\SetWatermarkVerCenter{170mm}
\SetWatermarkColor{darkgray}
\SetWatermarkText{Late-Breaking / Demo Session Extended Abstract, ISMIR 2024 Conference}

\def\lbd{}	

\title{Audio Atlas: Visualizing and Exploring Audio Datasets}

\multauthor
{Luca A. Lanzend\"orfer \hspace{0.9cm} Florian Gr\"otschla \hspace{0.9cm} Uzeyir Valizada \hspace{0.9cm} Roger Wattenhofer} { 
ETH Zurich\\
{\tt\small \{lanzendoerfer, fgroetschla, uvalizada, wattenhofer\}@ethz.ch}
}

\def\authorname{L. A. Lanzend\"orfer, F. Gr\"otschla, U. Valizada, and R. Wattenhofer}

\usepackage[bookmarks=false,pdfauthor={\authorname},pdfsubject={\papersubject},hidelinks]{hyperref}

\begin{document}

\maketitle
\begin{abstract}
We introduce Audio Atlas, an interactive web application for visualizing audio data using text-audio embeddings. Audio Atlas is designed to facilitate the exploration and analysis of audio datasets using a contrastive embedding model and a vector database for efficient data management and semantic search. The system maps audio embeddings into a two-dimensional space and leverages DeepScatter for dynamic visualization. Designed for extensibility, Audio Atlas allows easy integration of new datasets, enabling users to better understand their audio data and identify both patterns and outliers. We open-source the codebase of Audio Atlas, and provide an initial implementation containing various audio and music datasets.\footnote{\url{https://github.com/ETH-DISCO/audio-atlas}}
\end{abstract}

\section{Introduction}

The increasing size of machine learning datasets presents significant challenges in data visualization and analysis. Traditional tools are often insufficient for effectively managing and interpreting unlabeled audio datasets at scale. Having the ability to visualize large-scale datasets is crucial in helping to understand the structure and patterns within the data. It allows users to quickly grasp relationships between variables, identify trends, and detect outliers or anomalies that may affect the performance of a machine learning model.

Although there have been various projects focusing on providing high-level insight into audio and music datasets, they do not allow users to visualize their own data or were not concerned with large-scale datasets~\cite{EveryNoise,MusicMap,bogdanov2019acousticbrainz,EternalJukebox,smilkov2016embedding}. As such, these tools are mostly unsuitable for machine learning projects.
To this end, we present Audio Atlas, an open-source interactive web application that helps users navigate audio and music datasets. Inspired by existing work in the image domain~\cite{grötschla2024aeyevisualizationtoolimage}, Audio Atlas can visualize any audio dataset, providing a responsive user interface even when displaying tens of millions of samples.

To obtain semantically meaningful embeddings, we use CLAP~\cite{wu2024largescale}, a contrastive neural network trained on audio-text pairs. This enables Audio Atlas to display audio data with meaningful clusters and facilitates effective semantic searches and content exploration without requiring any audio metadata. 

\begin{figure}[t]
\centering
\includegraphics[width=\columnwidth]{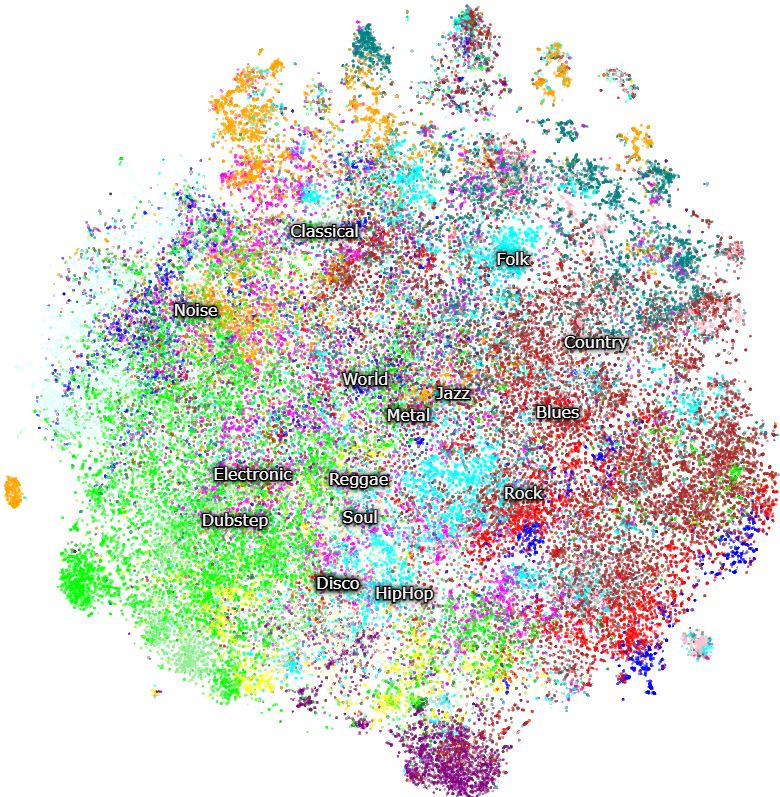}
\caption{Audio Atlas view of the FMA dataset~\cite{defferrard2017fmadatasetmusicanalysis}. A cluster can be highlighted by clicking on the label.}
\label{fig:graph_highlight}
\end{figure}

\section{Audio Atlas}

Audio Atlas is a visualization tool designed to help users interact with audio data through an intuitive and dynamic web interface. The application leverages the Contrastive Language-Audio Pretraining (CLAP)~\cite{wu2024largescale} model to generate embeddings that fuse audio and text into a shared vector space. These embeddings are then projected onto a two-dimensional plane using t-SNE~\cite{JMLR:v9:vandermaaten08a}, and are visualized as a point cloud. We use Milvus~\cite{10.1145/3448016.3457550} to store the CLAP embeddings. Milvus is a high-performance open-source vector database which efficiently manages embeddings and enables nearest-neighbor lookup for semantic searches with both text and audio snippets. 

\begin{figure}[t]
    \centering
    \includegraphics[trim={0 3cm 0 0}, clip, width=\linewidth]{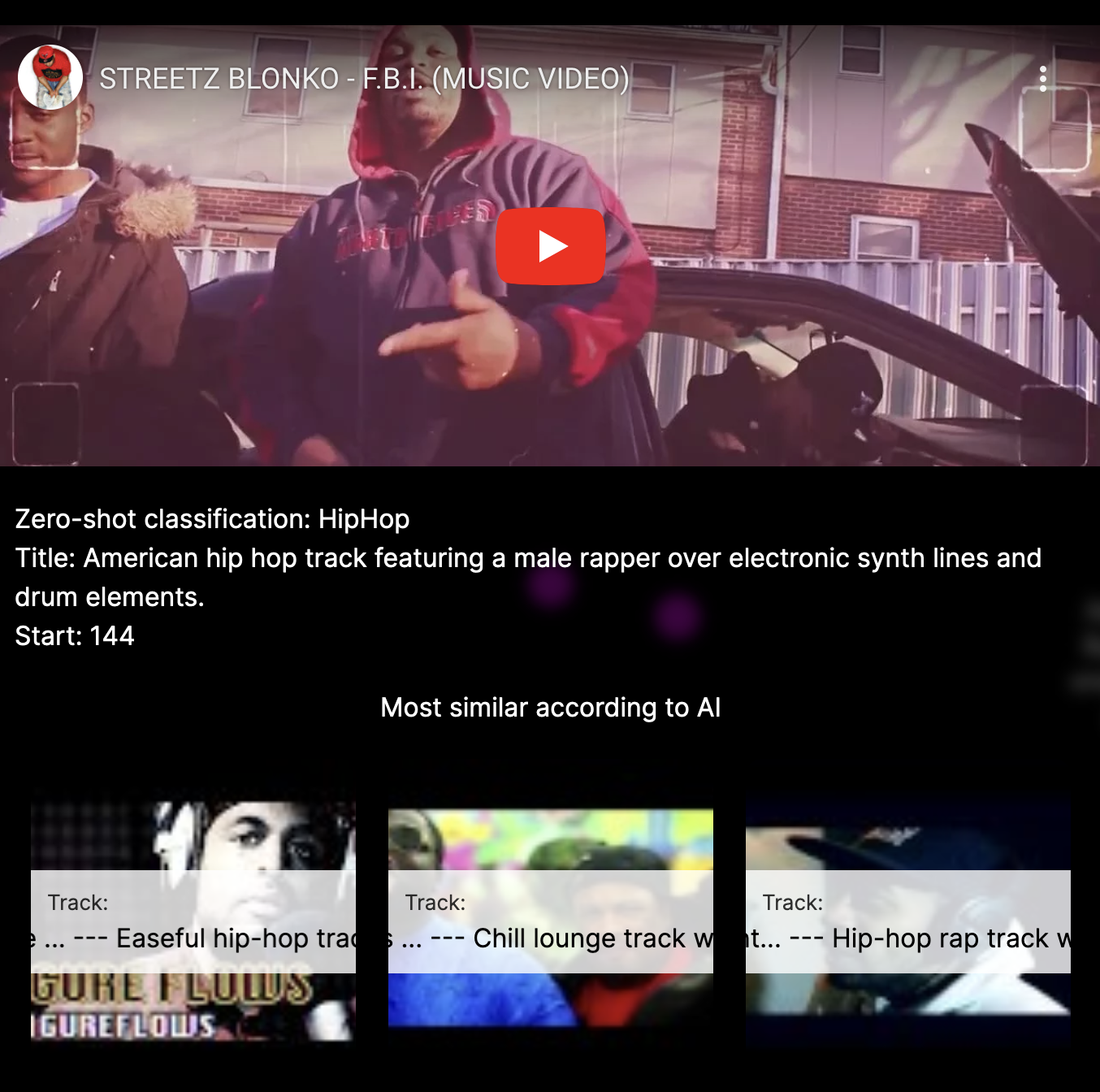}
    \caption{View when clicking an audio sample or using the search bar. Datasets containing links to YouTube are embedded to allow playback. The grid below shows the most similar results according to the CLAP embedding space.}
    \label{fig:datapoint}
\end{figure}

Audio Atlas enables users to perform zero-shot classification on their audio data. Zero-shot classification categorizes datasets without prior explicit training on specific classes, which is particularly useful when labeled data is scarce or missing entirely. 
The CLAP embeddings are used for zero-shot classification with a user-definable list of classes. The visualization highlights the classes with different colors, and a click on the label highlights only the selected class.
The frontend is powered by DeepScatter~\cite{DeepScatter}, built on top of WebGL~\cite{WebGL} and React~\cite{ReactJS}, and renders the visualizations interactively, allowing users to explore audio datasets by navigating through clusters, performing semantic searches, and listening to the audio snippets. A click on a datapoint opens a detailed view with more information provided in the dataset, such as classes, labels, and descriptions, as well as a list of closest neighbors in the embedding space sorted by similarity (cf.~Fig.~\ref{fig:datapoint}). Our initial implementation provides access to MusicCaps~\cite{agostinelli2023musiclm}, YT8M-MTC~\cite{mckee2023language}, VCTK~\cite{Yamagishi2019CSTRVC}, ESC-50~\cite{piczak2015dataset}, MTG-Jamendo~\cite{bogdanov2019mtg}, and FMA~\cite{defferrard2017fmadatasetmusicanalysis}. Additionally, Audio Atlas remains responsive on large-scale datasets~\cite{lanzendörfer2023disco10m}.

    Furthermore, Audio Atlas can be used for semantic search. Users can search for audio using text or audio, with the provided search bar in Audio Atlas. The modality provided in the search bar (text or audio) is converted into an embedding vector using a contrastive model. The nearest neighbors of this embedding vector are computed using Annoy, an approximate nearest neighbor search framework. This semantic search enables users to find and filter audio using descriptive text, which historically has only been possible if the audio data contained rich annotations. Using contrastive models, we can perform a semantic search on disjoint modalities. While using audio to search for audio has existed previously, using a contrastive approach allows us to search on the basis of semantic meaning instead of the similarity of extracted audio features or waveform similarity. 
    
\section{Use Cases}

We demonstrate a series of practical applications of Audio Atlas, showing its utility in audio analysis through dimensionality reduction and search techniques.

Audio Atlas makes it easy to browse an audio dataset using text and audio queries. We can therefore also qualitatively assess the classification capabilities of the contrastive embedding model. Furthermore, we can explore the dataset with descriptive queries instead of searching for specific tracks using their titles. 

\begin{figure}[t]
    \centering
    \includegraphics[width=0.89\linewidth]{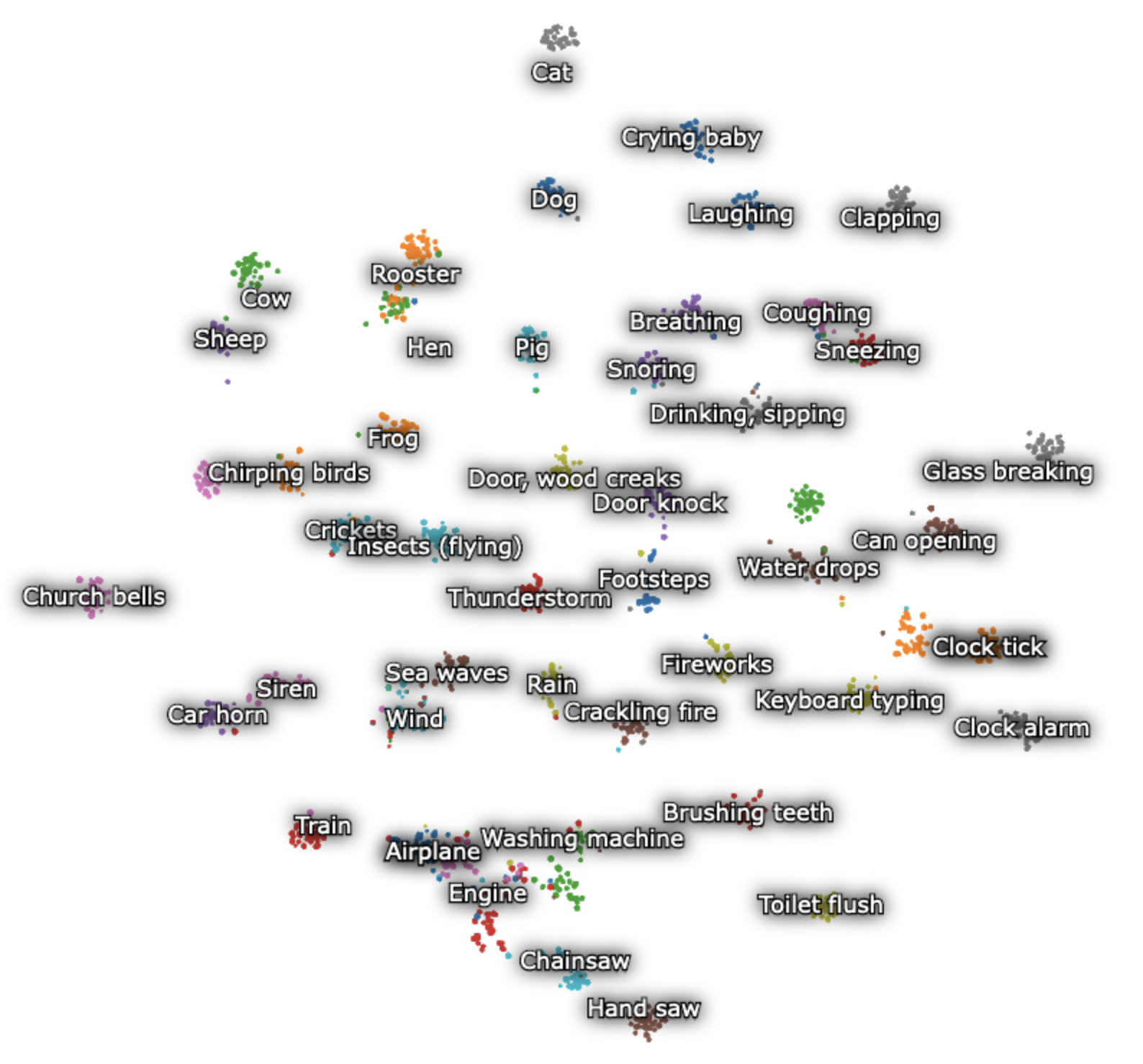}
    \caption{Audio Atlas visualization of ESC-50 dataset. The clusters of the various classes found in the Environmental Sound Classification dataset are clearly visible.}
    \label{fig:esc50tsne}
\end{figure}

Additionally, users can further explore datasets using the upload function with audio queries. By doing so, users are able to find the results that are most similar to their audio file in the dataset according to the similarity of the embedding space, as well as finding the most similar result's label and embedding location. This helps users understand the neighborhood of their query audio file.

In Figure~\ref{fig:esc50tsne}, we use the ESC-50 classes to classify the dataset using zero-shot classification. In this way, we can visualize the semantic meanings of the clusters. Moreover, we can see that the t-SNE projection of the CLAP embeddings for the ESC-50 dataset has clustered all classes into local pockets. This clustering helps us explore specific datasets, in addition to understanding what the CLAP model has learned. We can easily apply Audio Atlas to other datasets with various labels to classify and explore the data. Furthermore, we believe Audio Atlas could be used as a novel way to browse music samples. For example, users could explore their unannotated audio libraries simply by describing the sound they are looking for.

In summary, by transforming audio into a visual representation, Audio Atlas gives users a new tool to explore large-scale audio datasets interactively. By making the codebase for Audio Atlas open-source we hope to advance the study of large-scale audio datasets as well as qualitatively assessing the performance of embedding models.

\bibliography{ISMIRtemplate}

\end{document}